\input harvmac.tex
\input epsf.tex
\parindent=0pt
\parskip=5pt

\hyphenation{satisfying}

\def\IR{{\hbox{{\rm I}\kern-.2em\hbox{\rm R}}}}
\def\IB{{\hbox{{\rm I}\kern-.2em\hbox{\rm B}}}}
\def\IN{{\hbox{{\rm I}\kern-.2em\hbox{\rm N}}}}
\def\IC{\,\,{\hbox{{\rm I}\kern-.59em\hbox{\bf C}}}}
\def\IZ{{\hbox{{\rm Z}\kern-.4em\hbox{\rm Z}}}}
\def\IP{{\hbox{{\rm I}\kern-.2em\hbox{\rm P}}}}
\def\IH{{\hbox{{\rm I}\kern-.4em\hbox{\rm H}}}}
\def\ID{{\hbox{{\rm I}\kern-.2em\hbox{\rm D}}}}
\def\II{{\hbox{\rm I}\kern-.2em\hbox{\rm I}}}

\noblackbox

\leftline{\epsfxsize1.0in\epsfbox{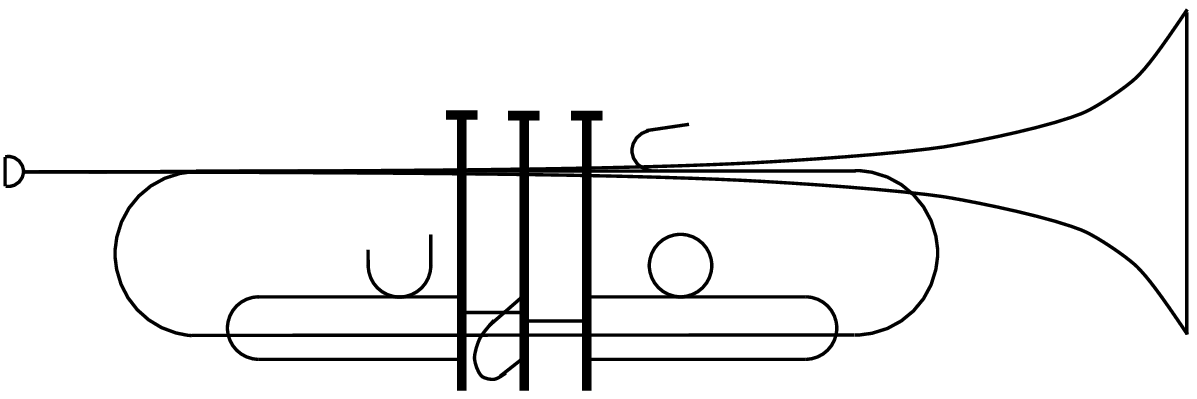}}
\vskip-1.0cm
\Title{\hbox{hep-th/9804201}}
{On the $(0,4)$  Conformal Field Theory of the Throat}

\centerline{\bf Clifford V. Johnson$^\flat$}

\bigskip
\bigskip

\vbox{\baselineskip12pt\centerline{\hbox{\it 
Department of Physics}}
\centerline{\hbox{\it University of California}}
\centerline{\hbox{\it Santa Barbara CA 93106, U.S.A.}}}

\footnote{}{\sl email: $^\flat${\tt cvj@itp.ucsb.edu}$\,\,$
On leave from the Dept. of
Physics and Astronomy, University of Kentucky, Lexington KY~40502,
U.S.A.}

\vskip1.0cm
\centerline{\bf Abstract}
\vskip0.7cm
\vbox{\narrower\baselineskip=12pt\noindent
In $SO(32)$ heterotic string theory, the space--time at the core of
$N$ coincident NS--fivebranes is an infinite  throat,
$\IR{\times}S^3$. As shown by Witten, the throat signals a singularity
in the usual heterotic string conformal field theory and a
non--perturbative $USp(2N)$ gauge group appears, due to the $N$ small
instantons at the fivebranes' core. Nevertheless, we look for some
trace of the non--perturbative physics in a description of the
heterotic string infinitely far down the throat. Our guide is a
D1--brane probing $N$ D5--branes in type~I, which yields a 1+1
dimensional $(0,4)$ supersymmetric model with ADHM data in its
couplings, as shown by Douglas. The neighbourhood of the classical
boundary of the hypermultiplet moduli space of the theory flows to an
exact conformal field theory description of the throat
theory. Ironically, the remnant of the non--perturbative symmetry is
indeed found in the conformal field theory, lurking in the structure
of the partition function, and encoded in a family of deformations of
the theory along flat directions. The deformations have an explicit
description using the flow from the type~I theory, and have a
hyperK\"ahler structure.  Similar results hold true for the analogous
$(4,4)$ supersymmetric situation in the type~IIB theory, as is evident
in the work of Diaconescu and Seiberg.}


\Date{31st March 1998} 
\baselineskip13pt
\lref\dbranes{J.~Dai, R.~G.~Leigh and J.~Polchinski, 
{\sl `New Connections Between String Theories'}, Mod.~Phys.~Lett.
{\bf A4} (1989) 2073\semi P.~Ho\u{r}ava, {\sl `Background Duality of
Open String Models'}, Phys. Lett. {\bf B231} (1989) 251\semi
R.~G.~Leigh, {\sl `Dirac--Born--Infeld Action from Dirichlet Sigma
Model'}, Mod.~Phys.~Lett. {\bf A4} (1989) 2767\semi J.~Polchinski,
{\sl `Combinatorics Of Boundaries in String Theory'}, Phys.~Rev.~D50
(1994) 6041, hep-th/9407031.}

\lref\berkoozi{M. Berkooz, R. G. Leigh, J. Polchinski, J. Schwarz, N. Seiberg 
and E. Witten, {\sl `Anomalies, Dualities, and Topology of D=6 N=1 Superstring
Vacua'}, Nucl. Phys. {\bf B475} (1996) 115, hep-th/9605184.}

\lref\nsfivebrane{A. Strominger,  {\sl `Heterotic Solitons'} Nucl. Phys. 
{\bf B343}, (1990) 167; 
{\it Erratum: ibid.}, {\bf 353} (1991) 565.}
\lref\robin{R. W. Allen, I. Jack and D. R. T. Jones, 
{\sl `Chiral Sigma Models and the Dilaton Beta Function'}, 
Z. Phys. C {\bf 41} 
(1988) 323.}

\lref\heteroticcosets{C. V. Johnson, {\sl `Heterotic Coset Models'},
 Mod. Phys. Lett. {\bf A10} (1995) 549, hep-th/9409062\semi
{\it ibid.,} {\sl `Exact Models of Extremal Dyonic 4D Black Hole Solutions
 of Heterotic String Theory'}, Phys. Rev. 
{\bf D50} (1994) 4032, hep-th/9403192.}

\lref\diacon{D.--E. Diaconescu and N. Seiberg, 
{\sl `The Coulomb Branch of $(4,4)$ Supersymmetric Field Theories in 
Two Dimensions'}, JHEP, {\bf 07}
 (1997) 001, hep-th/9707158.}
\lref\mikejoeandy{M. R. Douglas, J. Polchinski, and A. Strominger, 
{\sl`Probing
 Five Dimensional Black Holes with D--branes'}, JHEP
12 (1997) 003, hep-th/9703031.}
\lref\callan{C. G. Callan, J.A. Harvey and A. Strominger, 
{\sl `Supersymmetric String Solitons'}, in
Trieste 1991, proceedings, ``String Theory and Quantum Gravity'',
hep-th/9112030.}
\lref\sjrey{S--J. Rey, in
 {\sl `Superstrings and Particle Theory: Proceedings'}, edited by
   L. Clavelli and B. Harms, (World Scientific, 1990) 291\semi
   S--J. Rey, {\sl `The Confining Phase of Superstrings and Axion Strings'},
 Phys. Rev. {\bf D43} (1991) 526\semi I. Antoniades,
   C. Bachas, J. Ellis and D. Nanopoulos, {\sl `Cosmological String
   Theories and Discrete Inflation'}, Phys. Lett.  {\bf B211} (1988)
   393\semi {\it ibid.,} {\sl `An Expanding Universe in String
   Theory'}, Nucl. Phys. {\bf 328} (1989) 117.}
\lref\robdilaton{R.~C.~Myers, {\sl `New Dimensions for Old Strings'}, 
Phys. Lett. {\bf B199} (1987) 371.}
\lref\ff{B. L. Feigin and D. B. Fuchs, Funct. Anal. Appl. {\bf 16} (1982)
 114, {\it ibid.}, {\bf 17} (1983) 241.}
\lref\gepner{D. Gepner, {\sl `Spacetime Supersymmetry in Compactified 
String Theory and Superconformal Models'}, Nucl. Phys. {\bf B296} (1988) 757.}

\lref\wznw{S. P. Novikov, Ups. Mat. Nauk. {\bf 37} (1982) 3\semi
E. Witten, {\sl `Non--Abelian Bosonization in Two Dimensions'},
Comm. Math. Phys. {\bf 92} (1984) 455.}
\lref\rohm{R. Rohm, {\sl `Anomalous Interactions for the Supersymmetric 
Non--linear Sigma Model in Two Dimensions'}, Phys. Rev. {\bf D32}
(1984) 2849.}
\lref\edjoe{J. Polchinksi and E. Witten,  {\sl 
`Evidence for Heterotic - Type I 
String Duality'}, Nucl. Phys. {\bf B460} (1996) 525, hep-th/9510169.}

\lref\edcomm{E. Witten, {\sl `Some Comments On String Dynamics'}, in the 
Proceedings of {\sl Strings 95}, USC, 1995, hep-th/9507121.}

\lref\ericjoe{E. G. Gimon and J. Polchinski, {\sl `Consistency
 Conditions of Orientifolds and D--Manifolds'}, Phys. Rev. {\bf D54} (1996) 
1667, hep-th/9601038.}
\lref\wittenadhm{E. Witten, {\sl `Sigma Models and the ADHM Construction of 
Instantons'}, J.~Geom.  Phys. {\bf 15} (1995) 215, hep-th/9410052.}
\lref\edsmall{E. Witten, {\sl `Small Instantons in String Theory'},  Nucl. 
Phys. {\bf B460} (1996) 541, hep-th/9511030.}
\lref\douglasii{M. R.  Douglas, {\sl `Gauge Fields and D--Branes'},  
hep-th/9604198.}

\lref\ADHM{M. F. Atiyah, V. Drinfeld, N. J. Hitchin and Y. I. Manin, {\sl
`Construction of Instantons'} Phys. Lett. {\bf A65} (1978) 185.}

\lref\italiansii{D. Anselmi, M. Bill\'o, P. Fr\'e, L. Giraradello and A. 
Zaffaroni, {\sl `ALE Manifolds and Conformal Field Theories'},  Int. J. 
Mod. Phys. {\bf A9} (1994) 3007,  hep-th/9304135.}
\lref\dnotes{J. Polchinski, S. Chaudhuri and C. V. Johnson, {\sl `Notes on 
D--Branes'}, hep-th/9602052.}

\lref\joetasi{J. Polchinski, `TASI Lectures on D-Branes', hep-th/9611050.}
\lref\hull{C. M.  Hull, {\sl `String--String Duality in Ten Dimensions'}, 
 Phys. Lett. {\bf B357} (1995) 545,  hep-th/9506194.}

\lref\cft{See for example the wonderful book
 {\sl `Conformal Field Theory'}, P. di Francesco, P.~Matthieu and
 D. S\'en\'echal, Springer, 1997.}
\lref\anatomy{C. V. Johnson, {\sl`Anatomy of a Duality'}, 
hep-th/9711082, to appear in Nucl. Phys. {\bf B}.}
\lref\ganor{O. J. Ganor and A. Hanany, 
{\sl `Small $E_8$ Instantons and Tensionless Non--critical Strings'},
Nucl. Phys. {\bf B474} (1996) 122, hep-th/9602120.}
\lref\lambertone{{\sl `Quantizing the (0,4) Supersymmetric ADHM Sigma 
Model'}, N. D. Lambert, Nucl. Phys. {\bf B460} (1996) 221,
hep-th/9508039.}
\lref\lamberttwo{{\sl `D--brane Bound States and the Generalised ADHM 
Construction'}, N. D. Lambert, Nucl. Phys. {\bf B519} (1998) 214,
hep-th/9707156.}


\newsec{\sl Fivebranes and Instantons}
In the type~I string theory, D5--branes are forced\ericjoe\ to travel
in pairs by the orientifold projection $\Omega$. This dynamical unit
carries an $SU(2)$ gauge symmetry on their $5{+}1$ dimensional
world--volume\edsmall. Consider a coincidence of $N$ of these
pairs. The gauge symmetry on the world volume is then $USp(2N)$. 

In the dual $SO(32)$ heterotic string theory, the $N$ brane pairs
become $N$ Neveu--Schwarz (NS) fivebranes with zero size instantons at
their core, and the metric, dilaton and $H$--field are\nsfivebrane:
\eqn\heterotic{\eqalign{ds_H^2
&=\left(-dt^2+\sum_{i=1}^{5}dy_i^2\right)
+\left(e^{2\Phi_0}+{N\alpha^\prime\over x^2}\right)
\left(dx^2+x^2d\Omega_3^2\right)\cr
e^{\Phi}&=\left(e^{2\Phi_0}+{N\alpha^\prime\over x^2}\right)^{{1\over2}},
\quad H=-\epsilon_{\mu\nu\lambda}^{\phantom{\mu\nu\lambda}\kappa}
\partial_\kappa\Phi.}}
where $t$ is time, the $y_i$'s are coordinates on the world--volume,
$d\Omega_3^2$ is the metric on a round unit three sphere, and $x$ is
the radial variable in the transverse space.

Near the core of the configuration ({\it i.e.,} near $x{=}0$) ---or
equivalently, for large $N$--- the metric may be written in terms of a
new radial coordinate\callan\ $\sigma{=}\sqrt{N\alpha^\prime}
\log(x/\sqrt{N\alpha^\prime})$:
\eqn\hetagain{\eqalign{ds_H^2&= 
-dt^2+\sum_{i=1}^{5}dy_i^2
+\left(d\sigma^2+N\alpha^\prime d\Omega_3^2\right)\cr
\Phi&=-{\sigma\over\sqrt{N\alpha^\prime}} + 
{\rm constant},\quad H{=}-N\alpha^\prime\epsilon_3,}}
which is $\IR^7{\times}S^3$, where  $\epsilon_3$ is the $S^3$ volume form. 

The solution is smooth everywhere, (in contrast to the type~I
situation where the three--sphere vanishes at finite proper
distance). The asymptotic flat spacetime has decoupled from the
solution and is infinitely far away in these coordinates.

The appearance of the infinite throat $\IR{\times}S^3$ in the
transverse geometry, with a negative linear dilaton\robdilaton\
signaling the runaway growth of the string coupling far down it, has
been of interest for some time. In the current language\edsmall, the
appearance of such a throat, which occurs only when the instantons are
of zero size, signals a singularity in the perturbative description of
the heterotic string. Non--perturbative effects become important in
this regime, and they have the distinction that they persist no matter
how small one can make the heterotic string coupling far away from the
branes: the strong coupling effects merely occur farther down the
throat, the mouth of which is at a point in $\IR^4$.

Luckily, we have a concise perturbative description of these heterotic
non--perturbative effects in the dual type~I string theory\edsmall: A
gauge symmetry $USp(2N)$ (from 5--5 strings) appears on the
world--volume of the dual D5--branes, to accompany the $SO(32)$ from
the D9--branes (from 9--9 strings), accompanied by hypermultiplets in
the $\bf(2N,32)$, (5--9 strings) and the antisymmetric tensor of
$USp(2N)$ (5--5 strings).

One might wonder, however, if there is still some way of capturing
some sign of this non--perturbative physics in a heterotic string
description. Necessarily, this description will have to be of the
physics down the throat, in order to be consistent with the previous
two paragraphs. As the region deep inside the throat is far away from
the outside, it is conceivable that this may be possible, with the
expectation that the transition between the description inside the
throat and the description outside {\sl must} lie outside heterotic
perturbation theory.

With this reasoning in mind, one is led to be optimistic that there is
a {\sl conformal field theory} description of the heterotic string down the
throat, and that it might somehow contain the $USp(2N)$
structure. There is a limit to the optimism, of course. There is
simply no way that we can expect to find a conformal field theory
description containing $USp(2N)$ as a {\sl gauge symmetry}, as the
central charge of an affine algebra based on that group rapidly
exceeds 26 as $N$ increases. Therefore, the $USp(2N)$ must appear in a
more subtle way.

\newsec{\sl Probing in Type~I}

Let us return to the type~I picture for a while, in order to gather
clues. There, the D1--brane represents an infinitely heavy heterotic
string soliton, which ultimately becomes the light fundamental
heterotic string in the strong coupling limit.  We will consider one
D1--brane. (Formulae relevant to a collection of $2M$ D1--brane probes
may be found in ref.\anatomy.)

We locate the D5--branes at a point in
$(x^6,x^7,x^8,x^9)$, and the D1--brane's world--volume will lie in
$(x^0,x^1)$. This arrangement breaks the Lorentz group up as follows:
\eqn\lorentz{ 
 SO(1,9) \supset SO(1,1)^{01}\times SO(4)^{2345}\times SO(4)^{6789},}
where the superscripts denote the sub--spacetimes in which the
surviving factors act. Following refs.\refs{\wittenadhm,\douglasii}, we
may label the worldsheet fields according to how they transform under
the covering group (which acts as a global symmetry of our final 1+1
dimensional model):
\eqn\cover{G=
[SU(2)^\prime\times \widetilde{SU(2)}^\prime]_{2345}\times
[SU(2)_R\times SU(2)_L]_{6789},} with doublet indices
$(A^\prime,\tilde{A}^\prime,A,Y)$, respectively. 

Accordingly, the 16 supercharges of the type~I string decompose under
\lorentz\ first (due to the D--strings) as
${\bf 16}{=}{\bf 8}_+{+}{\bf 8}_-$ where $\pm$ subscripts denote a
chirality with respect to $SO(1,1)$, and furthermore (due to the
D5--branes) each $\bf 8$ decomposes into a pair of $\bf 4$'s of the
$SO(4)$'s.  The orientation projection $\Omega$ will later pick out
one of these spinors to carry the supersymmetry on the D1--brane
world--volume, giving a $1{+}1$ dimensional system with $(0,4)$
supersymmetry.

The spectrum of massless fields in the model produces a family of
fields on the world--volume, with coordinates $(x^0,x^1)$. 
The supersymmetry algebra is of the form\wittenadhm:
\eqn\susiealgebra{\{Q^{AA^\prime},Q^{BB^\prime}\}
=\epsilon^{AB}\epsilon^{A^\prime B^\prime} P_-,}
where $P_-{\equiv}-i\partial/\partial\sigma^-$, where
$\sigma^\pm{=}(x^0\pm x^1)/2$. Here, $\epsilon^{AB}$ and
$\epsilon^{A^\prime B^\prime}$ are the antisymmetric tensors of
$SU(2)$ and $SU(2)^\prime$, respectively, $A,B,A^\prime$ and
$B^\prime$ being doublet indices.

The various fields which arise in the model from the 1--1, 1--5 and
1--9 sector may be obtained using the standard D--brane calculus, 
reviewed in refs.\refs{\dnotes.\joetasi}. For the specific case here,
the fields are already worked out in ref.\douglasii, where it was
noticed that the model is isomorphic to the theory constructed in
ref.\wittenadhm.

The fields are as follows: In the 1--1 sector we get a family of
four--component scalars $b^{A^\prime\tilde{A}^\prime}$ which
parametrize the motion of the D1--brane inside the D5--branes. There
is also a hypermultiplet field $b^{AY}(x^0,x^1)$, parametrizing the
motion of the D1--brane transverse to the D5--branes.

The superpartners consist of the four component fermion
$\psi_-^{A\tilde{A}^\prime}$, the right--moving superpartner of the
four component scalar field $b^{A^\prime\tilde{A}^\prime}$, and
$\psi_-^{A^\prime Y}$) is the right--moving superpartner of
$b^{AY}$. The supersymmetry transformations are:
\eqn\susie{\delta b^{A^\prime\tilde{A}^\prime}
=i\epsilon_{AB}\eta^{A^\prime A}_{+} 
\psi_-^{B\tilde{A}^\prime}
\,\,\,{\rm and}\,\,\,
\delta b^{AY}=i\epsilon_{A^\prime B^\prime}\eta^{AA^\prime}_+
\psi_-^{B^\prime Y}.}

From the 1--9 sector, we have left--moving fermions $\lambda_+$ in the
$\bf{32}$, coming from the fact that the 1--9 strings end on the
D9--branes which carry a gauge symmetry $SO(32)$. These are the
current algebra fermions of the heterotic string\edjoe. We denote a
component by $\lambda_+^M$, where $M$ is a D9--brane index, whenever
we need to explicitly show the index structure under the global
symmetry of the D9--brane gauge group.

The 1--5 string excitations give a
supermultiplet in the $\bf{2N}$ of the D5--branes' $USp(2N)$ gauge group, 
with components~$(\phi^{A^\prime
m},\chi_-^{Am})$:
\eqn\susiei{\delta\phi^{A^\prime m}=i\epsilon_{AB}\eta^{A^\prime A}_{+}
\chi_-^{Bm}.} 
Here, $m$ is a D5--brane group theory index. (Note that because
$\Omega$ forces the D5--branes to be paired, $m$ is even, and so
$\phi$ naturally has components in multiples of four, like a
hypermultiplet. Actually, it is a ``twisted''
hypermultiplet.)  Also, (with components $\chi_+^{Ym}$),
there are left moving fermions $\chi_+^{Y}$ transforming in the
$\bf{2N}$.

The 9--5 and 5--5 hypermultiplets of the $5{+}1$ dimensional theory
appear as couplings in the 1+1 dimensional theory.  From the 5--5
sector there is the antisymmetric of $USp(2N)$ denoted
$X^{AY}_{mn}$. Meanwhile, the 9--5 sector produces a $\bf{(32,2N)}$,
denoted $h^{Am}_M$, with $m$ and $M$ showing off its choices in D5--
and D9--brane group theory.

We can now display the supersymmetry transformation relating them to
the left moving fields:
\eqn\susieii{\delta\lambda_+^M=\eta_+^{AA^\prime} C^M_{AA^\prime}
\,\,\,{\rm and}\,\,\,
\delta\chi_+^{Ym}=\eta_+^{AA^\prime} C^{Ym}_{AA^\prime},}
\eqn\cform{{\rm where}\,\quad C^M_{AA^\prime}=h_{A}^{Mm}\phi_{A^\prime m},
\,\,\,{\rm and}\,\,\,
C^{Ym}_{AA^\prime}=\phi_{A^\prime}^n(X^{Ym}_{An}-b_A^{Y}\delta_{n}^m).}

These precise transformations allow us to write the non--trivial part
of the $(0,4)$ supersymmetric 1+1 dimensional Lagrangian containing
the Yukawa couplings and the potential:

\eqn\lagrange{\eqalign{{\cal L}_{\rm tot}={\cal L}_{\rm kinetic}
-{i\over4}\int d^2\!\sigma\biggl[ &\lambda_+^M 
\left(
\epsilon^{BD}{\partial C^M_{BB^\prime} \over\partial b^{DY}}
\psi_-^{B^\prime Y}+
\epsilon^{B^\prime D^\prime}{\partial C^M_{B B^\prime}\over 
\partial\phi^{D^\prime m}}
\chi_-^{Bm}
\right)\biggr.\cr
\biggl.+&\chi_+^{Ym} 
\left(
\epsilon^{BD}{\partial C^{Ym}_{BB^\prime} \over\partial b^{DY}}
\psi_-^{B^\prime Y}+
\epsilon^{B^\prime D^\prime}{\partial C^{Ym}_{B B^\prime}\over 
\partial\phi^{D^\prime m}}
\chi_-^{Bm}
\right)\biggr.\cr
\biggl.+&{1\over2}
\epsilon^{AB}\epsilon^{A^\prime B^\prime}\left(C^M_{AA^\prime}
C^M_{BB^\prime}+C^{Ym}_{AA^\prime}C^{Ym}_{BB^\prime}\right)\biggr].
}}
This was derived in ref.\wittenadhm\ as the most general $(0,4)$
supersymmetric Lagrangian with these types of multiplets, providing
that the $C$ satisfy the condition:
\eqn\hyperkahler{C^M_{AA^\prime}
C^M_{BB^\prime}+C^{Ym}_{AA^\prime}C^{Ym}_{BB^\prime}+ C^M_{BA^\prime}
C^M_{AB^\prime}+C^{Ym}_{BA^\prime}C^{Ym}_{AB^\prime}=0,} which they
do\douglasii. ${\cal L}_{\rm kinetic}$ contains the usual kinetic terms
for all of the fields, and the required terms which complete them
into gauge invariant terms.

This theory is very
interesting. The structure of (part of) the moduli space of  vacua
$b^{AY}$ was shown in ref.\wittenadhm\ (see also ref.\douglasii\ for
the full translation into D1--brane probe language) to be equivalent
to specifying the data shown by Atiyah, Hitchin, Drinfeld and
Manin\ADHM\ required for the full specification of instantons. This
proved that D5--branes were indeed instantons of the D9--brane gauge
theory, and that under duality, they were indeed related to the NS
fivebranes which are instantons of the heterotic gauge group.

As noted in
ref.\wittenadhm, the potential is of the form
\eqn\pot{V=(h^2+(X-b)^2)\phi^2.} 
The branch of moduli space focused on in ref.\wittenadhm\ is the
sector parameterised by the expectation values of the 1--1
hypermultiplet fields $b^{AY}$, and the couplings $h^{Am}_{M},
X^{AY}_{mn}$.  The 1--5 fields $\phi$ get masses according to the form
of the potential in equation. \pot, and  they may be
integrated out. The Yukawa couplings of the fermions $\lambda_+,
\chi_+$ generate a minimal coupling (in their kinetic terms) to a
background gauge potential. The gauge connection is of a form set by
the hypermultiplet parameters $X,h$. It is self--dual because of the
$(0,4)$ condition
\hyperkahler, the $X$ control the positions and orientation of the
$N$ instantons while $h$ determines their size.

Our interest here is in the small instanton limit with the $2N$
D5--branes coincident, so we initially set the values of $X$ and $h$
to  zero. Now the potential is of the form $b^2\phi^2$, and
another branch of moduli space, parameterized by
$b{=}0,\phi{\neq}0$, seems to opens up. This classical analysis seems
to tell us that the  branches touch each other at $\phi{=}b{=}0$,
but an analysis of the global symmetries of the two branches suggests
that the origins of the $\phi$ branches are infinitely far away: The
$spin(4)$ ---the covering group of the $SO(4)_{2345}$--- acts on the
bosons $\phi^{A^\prime m}$ of the $\phi$ branch, but only on the
fermions $\psi^{A^\prime Y}_-$ of the $b^{AY}$ branch. Meanwhile, the
$SU(2)_R$ acts on the bosons $b^{AY}$ of the `$b$ branch' but only on
the fermions $\chi_{\pm}^{Am}$ of the `$\phi$ branch'. The two branches
of moduli space have different symmetries and therefore should not
touch each other, and the quantum theory should reveal this.

In two dimensions, the phrase ``moduli space of vacua'' is used with
the understanding that wavefunctions can (and do) actually spread all
over the moduli space. Therefore, for consistency with the previous
paragraph, there must be an {\sl infinite} wormhole separating the
two\edcomm.

We can proceed further by finding the metric on the $b$ branch.
General considerations give at one loop (it is not enough to do a tree
level calculation on the hypermultiplet branch with only $(0,4)$
supersymmetry) gives\foot{See ref.\refs{\mikejoeandy}. See also
refs.\refs{\lambertone,\lamberttwo}\ for analysis of the $b$--branch
and the $\phi$--branch, respectively, using anomaly arguments.}:

\eqn\metric{ds^2=\left({\rm C}+
 {N\alpha^\prime\over x^2}\right) (dx^2+x^2d\Omega_3^2), \quad {\rm
and}\quad H=-N\alpha^\prime\epsilon_3,} with the obvious coordinates
on the four--space parameterized by $b^{AY}$. The constant C is the
result of the tree level calculation, giving the hyperK\"ahler space
$\IR^4$, and the ${N/ x^2}$ results from letting the $N$ 1--5
hypermultiplet fields run around a loop. Strictly speaking, we have
allowed the $X$ fields to be merely {\sl near} zero to generate the
appropriate cubic couplings implied by this result. This is equivalent
to letting the D5--branes have the freedom to separate
slightly. Furthermore, in the expression above, we have worked with
$N(2N{-}1)$ equal masses $m_i$ for the 1--5 fields $\phi$,
corresponding to $N$ coincident D5--branes. In the case where we allow
such terms to be different ({\it i.e.,} use up the full family of
values for the $X^{AY}_{mn}$), the prefactor in the metric becomes
${\rm C}+\sum_{i=1}^{N}|x-m_i|^{-2},$ a fact which will be important
later.

The constant ${\rm C}$ in the metric is actually proportional to
$1/g_{YM}^2\sim1/g_I^2\sim g_h^2{=}e^{2\Phi_0}$.

After a change of variables to $\sigma=\sqrt{N\alpha^\prime}\log(
x/\sqrt{N\alpha^\prime})$ ({\it i.e.,} concentrating attention near the
origin of the $b$ branch) we see that we develop the anticipated
infinite throat geometry. We have recovered the fact that the origin
of the $\phi$ branch is infinitely far away.  In the stringy strong
coupling limit, this model flows to an infra--red fixed point and the
model is appropriate to the world--sheet theory of the heterotic
string.

(It is worth noting here that this result also applies to the more
general $(0,4)$ models derived in ref.\anatomy, in the context of type~I
D1--branes on an ALE space: The results of ref.\berkoozi\ showed that
even in the absence of D5--branes, the type~I model of ref.\ericjoe\
still contains instantons hidden in the ALE spaces. This was confirmed
directly with the probe models of ref.\anatomy, as it was shown that
there are additional fields in the model coming from 1--9 strings
`trapped' at the fixed points and therefore mimicking the behaviour of
1--5 strings. The resulting fields are therefore analogues of the
$\phi$ fields discussed here. Even in the absence of D5--branes
therefore, the resulting metric on the moduli space would see the
throat--type core, but in that situation connected to an external
spacetime with a $\IZ_2$ action at infinity.)

\newsec{\sl The Conformal Field Theory of the Throat}
In the stringy strong coupling
limit, this model flows to an infra--red fixed point.

The form of the throat geometry, with the linear dilaton is very
suggestive of an exact conformal field theory description based on
affine $SU(2)_{N}$ (a WZW model\wznw) times a Feigen--Fuchs\ff\ theory
with screening charge $(N\alpha^\prime)^{-1/2}$. This has been
discussed many times before, particularly for the (4,4) supersymmetric
case in the context of type II string theory\refs{\callan,\sjrey}.

We would like to try to understand more details about this theory, as
it is the conformal field theory description of the throat which we hoped to
be led to earlier in the discussion.

\medskip

{\sl 3.1 Basic Content}

In the case in hand, we need to find a description in terms of a
$(0,4)$ supersymmetric conformal field theory. We have two clues from
the heterotic side:
\item\item{\it (i)} The
right moving sector must be supersymmetric, with $c_R{=}15$, and so we
can expect that the theory on the right will match the type II
case. 
\item\item{\it (ii)} The full $SO(32)$ gauge 
symmetry must be present, as in the small instanton limit, it is
restored.  Therefore, the left moving sector must contain an affine
$SO(32)_1$, which has $c{=}16$. (This is consistent with the fact that
there is a left--moving $SO(32)$ current algebra carried by the
$\lambda_+$ before the flow: It survives the limit.) The remaining
part of the left--moving sector must supply a central charge 10 for a
total of $c_L{=}26$.

Let us remind ourselves of how the right--moving supersymmetric sector
works. The six world--volume directions each supply a free boson and
majorana fermion, giving a central charge
$6{\times}{3\over2}{=}9$. The $SU(2)_N$ theory is
supersymmetried\rohm\ by adding three majorana fermions with values in
the Lie algebra. They can be made into free fermions by a chiral
rotation. This is at the cost\robin\ of shifting the level by
$N{\to}N{-}2$.  The central charge of the theory is therefore 9 from
the world--volume, $3{-}{6\over N}{+}{3\over2}$ from the
supersymmetric affine $SU(2)_N$ theory, toghether with $1{+}{6\over
N}{+}{1\over 2}$ from the Fiegen--Fuchs theory, where we have added a
fermion for supersymmetry. This gives $c_R{=}15$.

For the left--moving sector, we take the $c{=}16$ affine $SO(32)_1$
theory and the $c{=}6$ contribution from the flat non--supersymmetric
world--volume directions. The remaining central charge of 4 comes from
taking only the bosonic part of the previous paragraph: $3{-}{6\over
N}{+}1{+}{6\over N}{=}4$, giving $c_L{=}26$, as required\foot{Note
that this is in contrast to what has been done in interesting other
work on the symmetric throat conformal field theory in the
literature\italiansii, to make heterotic backgrounds. There, the full
supersymmetric $c_L{=}c_R{=}6$ theory was used as a building block to
construct $(0,4)$ models with some of the heterotic gauge symmetry
broken, effectively embedding the spin connection into the gauge group
in the standard way.}.

It is important to note that the final central charges are independent
of $N$ (of course) in a way which requires a balance between the
dilaton (Fiegen--Fuchs) sector, the WZW model, and the fermions.

So it appears that we can construct the required heterotic $(0,4)$
conformal field theory as the theory on the hypermultiplet branch,
representing the infinite throat physics.

We have one more extremely vital thing to specify, however. We need to
establish that we can indeed construct a modular invariant partition
function for the theory, and verify that it is the correct one implied
by the flow from the type~I theory. Is this possible?

Let us focus on the non--trivial $SU(2)_{N}$ part, for a moment.  If
this were the type II theory, we would have the same supersymmetric
structures on the left and the right, and the obvious ``A--type''
diagonal combination of the characters would work perfectly, as well
as other non--diagonal ``$D$--'' and ``$E$--type''
combinations. Overall, we are of course, forced to have a
non--diagonal combination in this heterotic case.

One set of modular invariants can be deduced as follows: Let us start
with with a $(4,4)$ model with A, D, or E modular invariants chosen
for the level $N{-}2$ WZW sector. Let us gather the eight majorana
fermions on each side into an affine $SO(8)_1$ current algebra. We can
use the diagonal modular invariant combination of those, for that
sector. Similarly for the bosonic Fiegen--fuchs theory. 

Now we can construct a heterotic modular invariant as follows: On the
left, take out the $SO(8)_1$ and replace it with an $SO(32)_1$. There
is a modular invariant combination of these two theories, after a
trivial exchange of the sign of the spinor and vector
representations\gepner. We now have $c_R{=}15, c_L{=}26$.

\medskip

{\sl 3.2 Constraints from Duality}

These are {\sl not} the conformal field theories implied by the flow
of the linear sigma model from type~I to heterotic. In particular, we
know that the conformal field theory has {\sl a family of
deformations} with a particular structure: This is encoded in the
freedom to generate  mass--difference terms $m_i$ for the $N$ 1--5
fields $\phi^{A^\prime}_m$, corresponding to different expectation
values of the $X^{AB}_{mn}$ hypermultiplets, allowing the D5--branes
to move apart. The flow from the linear sigma model implies that the
final conformal field theory must contain these deformations.

That the freedom to give vacuum expectation values to the
$X^{AB}_{mn}$ fields (generating mass differences $m_i$ for the
$\phi^{A^\prime}_m$'s) in the gauge theory corresponds to a family of
deformations of our conformal field theory is straightforward to
see. For example, adding a single mass difference $m_i$ will produce
the geometry:
\eqn\geometry{ds^2=\left({\rm C}+{1\over x-m}+
{N-1\over x}\right)(dx^2+x^2d\Omega_3^2).}  So we have two throats
after the flow of the model to the fixed point, one at $x{=}0$ and the
other at $x{=}m$. The first will contain the affine $SU(2)$ at
level~1, while the other has level $N{-}1$. Had we not generated $m_i$,
we would have had a single throat containing affine $SU(2)$ and level
$N$. However, recalling the discussion near the beginning of this
section, all of these conformal field theories have $(c_R{=}15,
c_L{=}26)$ due to the interplay of supersymmetry and the linear
dilaton. In other words, from the point of view of (say) the first
conformal field theory ($N$), we have added an operator which produced
a genuine deformation, as it preserved the overall central charge. The
resulting conformal field theories ($N{-}1$ and $1$) have now flown
infinitely far apart.

This structure is all inherited from the fact that we can move branes
apart in the dual theory, a freedom parametrized by the $USp(2N)$
antisymmetric tensor quartet $X^{AY}_{mn}$, the mass differences of
the 1--5 fields $\phi^{A^\prime}_m$. Furthermore, as this costs no
energy (the D5--branes are BPS states) we see that our family of
conformal field theory deformations define true flat directions in
this space of $(0,4)$ conformal field theories. (Note also that this
family of deformations has a hyperK\"ahler structure, inherited from
the parent instanton moduli space.)

This realization of a family of deformations of the $(0,4)$ conformal field
theory is amusing and satisfyingly explicit.

(Incidentally, there is another reason why the above choices cannot be
relevant to our problem: We could have replaced the left $SO(8)_1$
with level 1 affine $E_8{\times}E_8$ and arrived at the same
solutions. This is of course reminding us of the other heterotic
string. However, we know from multiple duality considerations that the
behaviour of small $E_8{\times}E_8$ instantons is very
different\refs{\edsmall,\ganor}, and not obtainable in this way. The
fact that the $E_8{\times}E_8$ current algebra can be inserted in such
a trivial way is further evidence that these modular invariants
proposed in the previous paragraph are not the solutions relevant to
this discussion, and should be regarded as artifacts.)

In summary, the linear sigma model implies the existence of a family of
deformations (but {\sl not} symmetries, of course) of the $(0,4)$ conformal
field theory of the throat, with a $USp(2N)$ structure. This data is
encoded in the conformal field theory in terms of certain restrictions
on how the various primary fields in the theory are constructed out of
the left and right--moving sectors. In other words, the $USp(2N)$ is
encoded in the required modular invariant mass matrix $M_{ij}$
appearing in the partition function:
\eqn\party{Z={\rm Tr}(q^{L_0-{C_L\over24}}
 {\bar q}^{{\bar L}_0-{C_R\over12}})
=\sum_{ij}M_{ij}\chi^i(q){\bar\chi}^j({\bar q}).}

We do not propose to embark on a search for an explicit form for this
$USp(2N)$ modular invariant partition function here, but merely stress
that its existence is ensured by duality\foot{It would be interesting
to find it explicitly, though. It may be possible to obtain a
description as a ``heterotic coset model''\heteroticcosets.}. Of
course, it is more standard to discuss modular invariants' isomorphic
structure to the Dynkin diagrams of simply laced Lie algebras. Here,
we are invoking an isomorphism to the non--simply laced algebra
$C_N$. This is not unreasonable. Recall\cft\ that the isomorphism is
between the possible indices which can appear in the sum \party\ and
the eigenvalues of the adjacency matrix of the Dynkin diagram, given
essentially by integers called the ``exponents''. The exponents of
$C_N$ are simply the odd integers $1,3,\ldots, 2N{-}1$, while for
$A_N$ it is the integers up to $1,2,\ldots, N$ and for $D_N$ it is
$1,3,\ldots,2N-3$, and $N{-}1$. We expect that the exponents of $C_N$
will appear in the sum instead of those of $D_N$ by a projection which
removes characters corresponding to the extra exponent $N{-}1$ and
replaces it with $2N{-}1$. So in this way only the exponents of $C_N$
will appear in the sum, and so it is not unthinkable to find $C_N$
characterizing a modular invariant\foot{In terms of Dynkin diagrams,
this is tantamount to projecting the diagram of $D_N$ so that the two
dovetailed roots at one end becomes the single long root at the end of
the $C_N$ diagram.}.

\newsec{ Closing Remarks}
We have found what we were looking for. 

$\bullet$ There {\sl is} a heterotic description of the physics of the
infinite throat in terms of a (0,4) conformal field theory, designed
as a heterotic variant of the exact conformal field theory presented
in the literature for the (4,4) type II case.  The conformal field
theory is IR limit of the linear sigma model description of the
apparent boundary of the $b^{AY}$ hypermultiplet branch of moduli
space of the theory on the world--volume of a D1--brane probing the
D5--branes of the type~I theory.

$\bullet$ There {\sl is} a remnant of the non--perturbative (from the
point of view of the usual heterotic variables) $USp(2N)$ 
symmetry living in this new description of the infinite throat. It
encodes the modular invariant combination of affine characters which
build the conformal field theory and is explicitly realized as an
important family of deformations of the theory along truly flat
directions. It did not appear as a gauge symmetry, as anticipated from
general considerations of consistency.

This picture is consistent with what happens in the type~IIB
case. There, we may also engage in a discussion involving D1--branes
and D5--branes, or fundamental strings and NS--fivebranes in the dual
type~IIB theory. In that case, the theory is $(4,4)$ supersymmetric,
and the situation was studied by Diaconescu and Seiberg\diacon. The
conformal field theory (arising this time from the Coulomb branch:
vector multiplet moduli space) is the usual symmetric $SU(2)_N$ WZW$+$
Feigen--Fuchs. The diagonal modular invariant $A_{N-1}$ encodes the
fact that there is an $SU(N)$ gauge symmetry on the D5--branes not
seen directly by the conformal field theory of the fundamental
strings, but by the D1--strings. In the case where there is an
orientifold O5--plane present (as distinct from an O9--plane), there
is an $SO(2N)$ gauge symmetry on the D5--branes, and the resulting
conformal field theory has the $D_{N-2}$ modular invariant.


\bigskip
\bigskip

\noindent
{\bf Acknowledgments:}

\noindent
CVJ was supported in part by family, friends and music. CVJ gratefully
acknowledges useful conversations with Robert C. Myers.  I am grateful
to Neil D. Lambert for making me aware of
refs.\refs{\lambertone,\lamberttwo}\ after an earlier version of this
manuscript appeared. This research was supported financially by NSF
grant \# NSFPHY97--22022.

\bigskip
\bigskip

\centerline{\epsfxsize1.0in\epsfbox{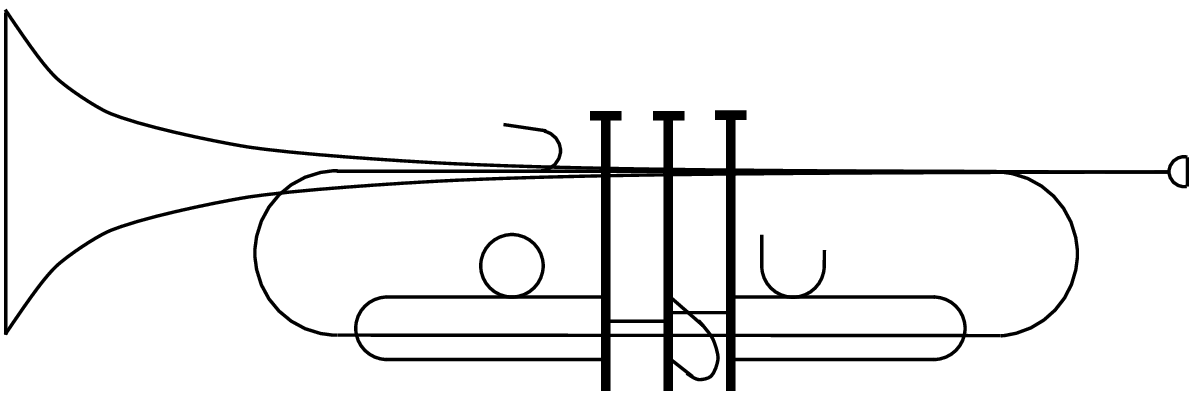}}

\listrefs

\bye